\def\annrev{{ARA\&A}}
\def\apj{{ApJ}}

\def\mnras{{MNRAS}}
\def\nat{{Nature}}

\def\vs{{vs.}}
\def\etal{{et al.}}
\def\farcm{\hbox{$.\mkern-4mu^\prime$}}

\def\kmsmpc {~km~s$^{-1}$~Mpc$^{-1}$}

\documentclass[cup5b]{caps}
\usepackage{graphicx}
\usepackage{amssymb}
\usepackage{ociwsymp2}
\HeadText{E. L. Wright}

\newcommand{\bc}{\begin{center}}    \newcommand{\ec}{\end{center}}
\newcommand{\bn}{\begin{enumerate}} \newcommand{\en}{\end{enumerate}}
\newcommand{\bi}{\begin{itemize}}   \newcommand{\ei}{\end{itemize}}
\newcommand{\be}{\begin{equation}}  \newcommand{\ee}{\end{equation}}
\newcommand{\bea}{\begin{eqnarray}} \newcommand{\eea}{\end{eqnarray}}

\begin{document}

\pagenumbering{arabic}


\author[]{EDWARD L. WRIGHT\\Department of Astronomy, University of California, Los Angeles}

\chapter{Theoretical Overview of Cosmic \\ Microwave Background Anisotropy}

\begin{abstract}
The theoretical basis for the prediction of anisotropies in the cosmic
microwave background is very well developed.  Very low amplitude
density and temperature perturbations produce small gravitational
effects, leading to an anisotropy that is a combination of temperature
fluctuations at the surface of last scattering and gravitational
redshifts both at last scattering and along the path to the
observer.  All of the primary anisotropy can be handled by linear
perturbation theory, which allows a very accurate calculation of the
predicted anisotropy from different models of the Universe.
\end{abstract}

\section{Introduction}

\begin{figure*}[t]
\includegraphics[width=1.0\columnwidth,angle=0,clip]{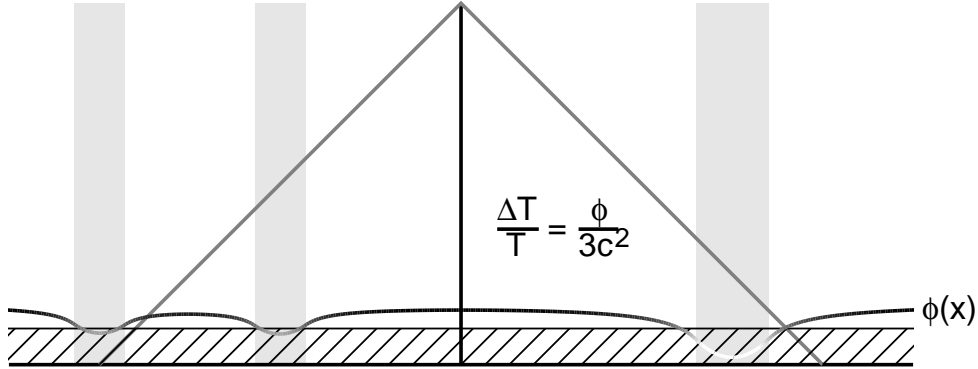}
\vskip 0pt \caption{
Sachs \& Wolfe (1967) predicted that density enhancements
would be cold spots in the CMB, as shown in this conformal spacetime diagram.
\label{fig:sax_wolf}}
\end{figure*}

The first predictions of the anisotropy of the cosmic microwave background
(CMB) were published shortly after the CMB was discovered by
Penzias \& Wilson (1965).
Sachs \& Wolfe (1967) calculated the anisotropies due to 
gravitational potential fluctuations produced by density perturbations 
(Figure \ref{fig:sax_wolf}).
Because the density perturbations are given by the second derivative
of the gravitational potential fluctuation in Poisson's equation, the
Sachs-Wolfe effect dominates the temperature fluctuations at large
scales or low spherical harmonic index $\ell$.
Sachs \& Wolfe predicted $\Delta T/T \approx 10^{-2}$ at large
scales.  This prediction, which failed by a factor of $10^3$, is based
on correct physics but incorrect input assumptions: prior to the
discovery of the CMB no one knew how uniform the Universe was on
large scales.

Silk (1968) computed the density perturbations needed at the 
recombination epoch at $z \approx 10^3$ in order to produce galaxies,
and predicted $\Delta T/T \approx 3 \times 10^{-4}$ on arcminute scales.
Silk (1967) calculated the damping  of waves that were partially
optically thick during recombination.  This process, known as ``Silk damping,''
greatly reduces the CMB anisotropy for small angular scales.

Observations by Conklin (1969) and then Henry (1971) showed that there was 
a dipole anisotropy in the CMB corresponding to the motion of the Solar
System with respect to the average velocity of the observable Universe.
There is a discussion of the dipole observations and their interpretation
in Peebles (1971) that is still valid today, except that what was then a
``tentative'' dipole is now known to better than 1\% accuracy, after
a string of improved measurements starting with Corey \& Wilkinson (1976)
and ending with the {\it COBE} DMR (Bennett \etal\ 1996).

Peebles \& Yu (1970) calculated the baryonic oscillations resulting
from interactions between photons and hydrogen in the early Universe,
and also independently introduced the Harrison-Zel'dovich spectrum
(Harrison 1970; Zel'dovich 1972).  Pebbles \& Yu predicted 
$\Delta T/T \approx 1.5 \times 10^{-4}$ on $1^\prime$ scales and
$\Delta T/T \approx 1.7 \times 10^{-3}$ on $7^\prime$ scales.

Wilson \& Silk (1981) further developed the theory of photon
and matter interaction by scattering and gravity, and predicted
$\Delta T/T = 100\;\mu$K for a single subtracted experiment
with a $7^\circ$ throw and with a $7^\circ$ beam like {\it COBE}.
Of course, when {\it COBE} was launched in 1989 it actually
observed a much smaller anisotropy.

These early predictions of a large anisotropy were greatly modified
by the addition of dark matter to the recipe for the cosmos.
Observational upper limits on small-scale anisotropies had
reached $\Delta T/T < 4 \times 10^{-5}$ on 1\farcm5 scales
(Uson \& Wilkinson 1982), which was considerably less than
the predictions from universes with just baryons and photons.
Peebles (1982) computed the anisotropy expected in a universe ``dominated
by massive, weakly interacting particles'' --- in other words cold
dark matter (CDM), although this paper predated the use of ``cold
dark matter.''

Bond \& Efstathiou (1987) calculated the correlation function
of the CMB anisotropy, $C(\theta)$, and also the angular
power spectrum, $C_\ell$, in the CDM cosmology.
This paper contains one of the first
plots showing $\ell(\ell+1)C_\ell$ \vs\ $\ell$, with peaks originally
called the ``Doppler'' peaks but more properly called ``acoustic peaks.''
This paper solved the Boltzmann equation describing the evolution
of the photon distribution functions.  Several authors developed
these ``Boltzmann codes,'' but the calculation of the angular power
spectrum up to high $\ell$ was very slow.
These codes described the conversion of inhomogeneity at the
last-scattering surface into anisotropy on the observed sky by
a set of differential equations evolving the coefficients of
a Legendre polynomial expansion of the radiation intensity.
Since the Universe is almost completely transparent after recombination,
a ray-tracing approach is much more efficient.
This great step in efficiency was implemented in the CMBFAST
code by Seljak \& Zaldarriaga (1996).

Hu \& Dodelson (2002) give a recent review of CMB anisotropies,
which includes a very good tutorial on the theory of $\Delta T/T$.

\section{Results}

A simple analysis of cosmological perturbations can be obtained from
a consideration of the Newtonian approximation to a homogeneous
and isotropic universe.  Consider a test particle at radius $R$
from an arbitrary center.  Because the model is homogeneous the choice
of center does not matter.  The evolution of the velocity of the test
particle is given by the energy equation
\be
\frac{\upsilon^2}{2} = E_{tot} + \frac{GM}{R}.
\label{eq:energy}
\ee
If the total energy $E_{tot}$ is positive, the Universe will expand forever
since $M$, the mass (plus energy) enclosed within $R$, is positive, $G$ is
positive, and $R$ is positive.
In the absence of a cosmological constant or ``dark energy,'' the 
expansion of the Universe will stop, leading to a recollapse if
$E_{tot}$ is negative.  But this simple connection between $E_{tot}$ and
the fate of the Universe is broken in the presence of a vacuum energy
density.
The mass $M$ is proportional to $R^3$ because the Universe is
homogeneous and the Hubble velocity $\upsilon$ is given by $\upsilon = HR$.  
Thus $E_{tot} \propto R^2$.

We can find the total energy by plugging
in the velocity $\upsilon_0 = H_0 R_0$ and the density $\rho_0$
in the Universe now.  This gives
\be
E_{tot} = \frac{(H_0 R_0c)^2}{2} -
 \frac{4\pi G \rho_0 R_0^2}{3} =
\frac{(H_0 R_0)^2}{2}
\left(1 - \frac{\rho_0}{\rho_{crit}}\right), 
\ee
with the critical density at time $t_0$ being
$\rho_{crit} = 3 H_0^2/(8\pi G)$.
We define the ratio of density to critical density as
$\Omega = \rho/\rho_{crit}$.  
This $\Omega$ includes all forms of matter and energy.  $\Omega_m$ will
be used to refer to the matter density.

From Equation \ref{eq:energy} we can compute the time variation of $\Omega$.
Let
\be
2 E_{tot} = \upsilon^2 - \frac{2 G M}{R} = H^2 R^2 - \frac{8\pi G \rho R^2}{3} =
\mbox{const}.
\ee
If we divide this equation by $8\pi G \rho R^2/3$ we get
\be
\frac{3 H^2}{8 \pi G \rho} - 1 = \frac{\mbox{const$^\prime$}}{\rho R^2}
= \Omega^{-1} - 1.
\ee
Thus $\Omega^{-1} - 1 \propto (\rho R^2)^{-1}$.
When $\rho$ declines with expansion at a rate faster than $R^{-2}$
then the deviation of $\Omega$ from unity grows with expansion.
This is the situation during the matter-dominated epoch with
$\rho \propto R^{-3}$, so $\Omega^{-1} - 1 \propto R$.  During
the radiation-dominated epoch $\rho \propto R^{-4}$, so
$\Omega^{-1} - 1 \propto R^2$.
For $\Omega_0$ to be within 0.9 and 1.1, $\Omega$ needed
to be between 0.999 and 1.001 at the epoch of recombination,
and within $10^{-15}$ of unity during nucleosynthesis.
This fine-tuning problem is an aspect of the ``flatness-oldness''
problem in cosmology.

Inflation produces such a huge expansion that quantum fluctuations on the
microscopic scale can grow to be larger than the observable Universe.
These perturbations can be the seeds of structure formation and also will
create the anisotropies seen by {\it COBE} for spherical harmonic indices
$\ell \geq 2$.  For perturbations that are larger than $\sim c_s t$
(or $\sim c_s/H$) we can ignore pressure gradients, since pressure gradients
produce sound waves that are not able to cross the perturbation in
a Hubble time.  In the absence of pressure gradients, the density perturbation
will evolve in the same way that a homogeneous universe does, and we can use
the equation
\be
\rho a^2 \left(\frac{1}{\Omega}-1\right) = \mbox{const}, 
\ee
the assumption that $\Omega \approx 1$ for early times,
and $\Delta\rho \ll  \rho$ as indicated by the smallness of the $\Delta T$'s
seen by {\it COBE}, to derive 
\be
-\rho a^2 \left(\frac{1}{\Omega}-1\right) \approx
\rho_{crit} a^2 \Delta\Omega \approx \Delta\rho a^2 = \mbox{const}.
\ee
Hence,
\be
\Delta\phi = \frac{G\Delta M}{R} = 
\frac{4\pi}{3}\; \frac{G \Delta\rho_0 (aL)^3}{aL} = 
\frac{1}{2}\;\frac{\Delta\rho_0}{\rho_{crit}}\; (H_0 L)^2,
\ee
where $L$ is the comoving size of the perturbation.  This is independent of 
the scale factor, so it does not change due to the expansion of the Universe.

During inflation (Guth 2003),
the Universe is approximately in a steady state with constant
$H$.  Thus, the magnitude of $\Delta\phi$ for perturbations with physical
scale $c/H$ will be the same for all times during the inflationary epoch.
But since this constant physical scale is $aL$, and the scale factor $a$ 
changes by more than 30 orders of magnitude during inflation, this means
that the magnitude of $\Delta\phi$ will be the same over 30 decades of
comoving scale $L$. Thus, we get a strong prediction that $\Delta\phi$
will be the same on all observable scales from $c/H_0$ down to
the scale that is no longer always larger than the sound speed horizon.
This means that
\be
\frac{\Delta\rho}{\rho} \propto L^{-2}, 
\ee
so the Universe becomes extremely homogeneous on large scales even though it
is quite inhomogeneous on small scales.

This behavior of $\Delta\phi$ being independent of scale is called
{\em equal power on all scales.}  It was originally predicted by 
Harrison (1970), Zel'dovich (1972),
and Peebles \& Yu (1970) based on a very simple argument:
there is no scale length provided by the early Universe, and thus the
perturbations should be {\em scale-free} ---  a power law.  Therefore
$\Delta\phi \propto L^m$.
The gravitational potential divided by $c^2$ is a component of the metric,
and if it gets comparable to unity then wild things happen.  If $m < 0$
then $\Delta\phi$ gets large for small $L$, and many black holes would form.
But we observe that this did not happen.
Therefore $m \geq 0$.  But if $m > 0$ then $\Delta\phi$ gets large on
large scales, and the Universe would be grossly inhomogeneous.
But we observe that this is not the case, so $m \leq 0$.  Combining both
results requires that $m = 0$, which is a {\em scale-invariant} perturbation
power spectrum.  This particular power-law power spectrum is called the
Harrison-Zel'dovich spectrum.
It was expected that the primordial perturbations
should follow a Harrison-Zel'dovich spectrum because all other answers were
wrong, but the inflationary scenario provides a good mechanism for producing
a Harrison-Zel'dovich spectrum.

\begin{figure*}[t]
\includegraphics[width=0.48\columnwidth,angle=0,clip]{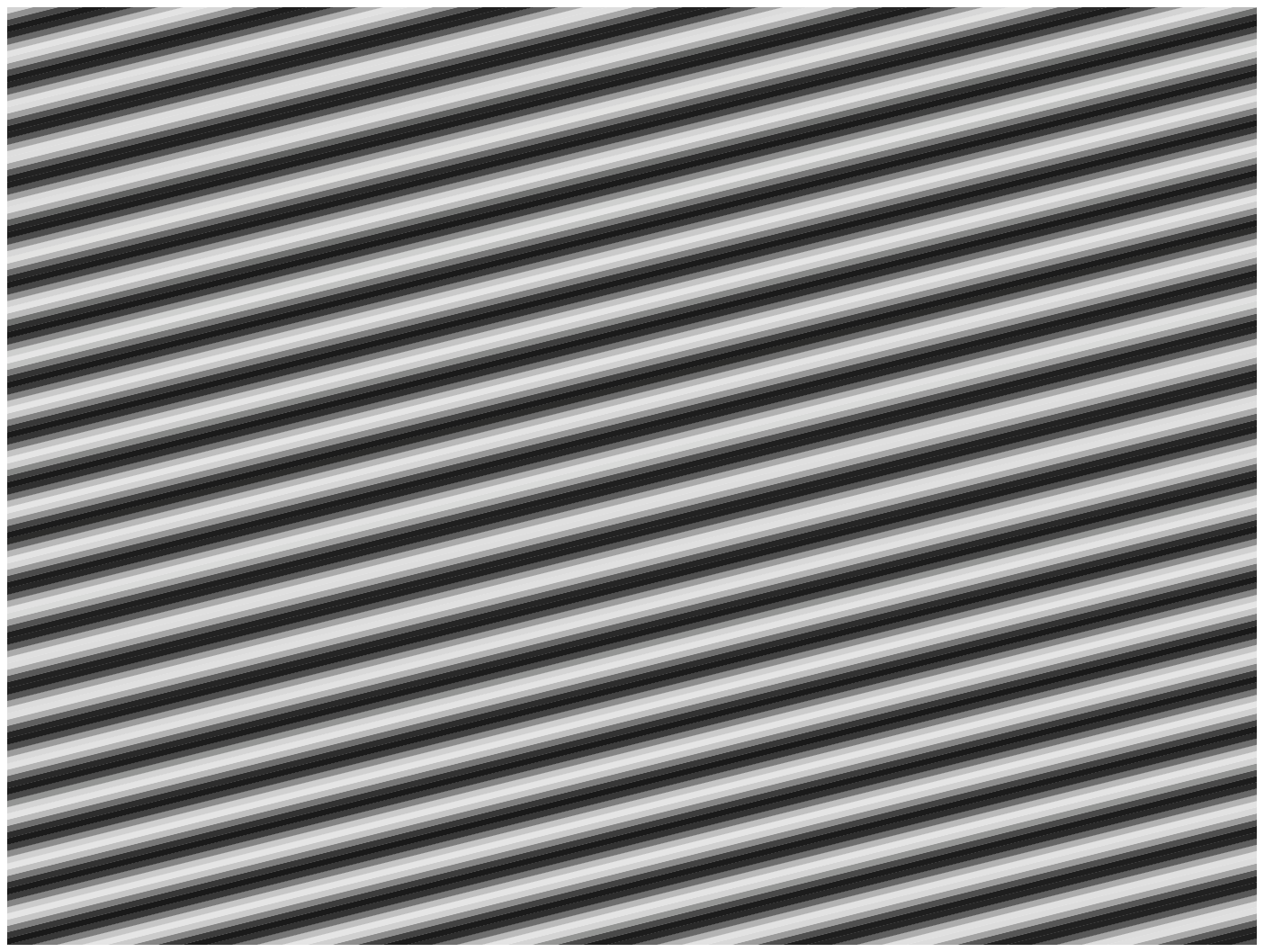} \hfil
\includegraphics[width=0.48\columnwidth,angle=0,clip]{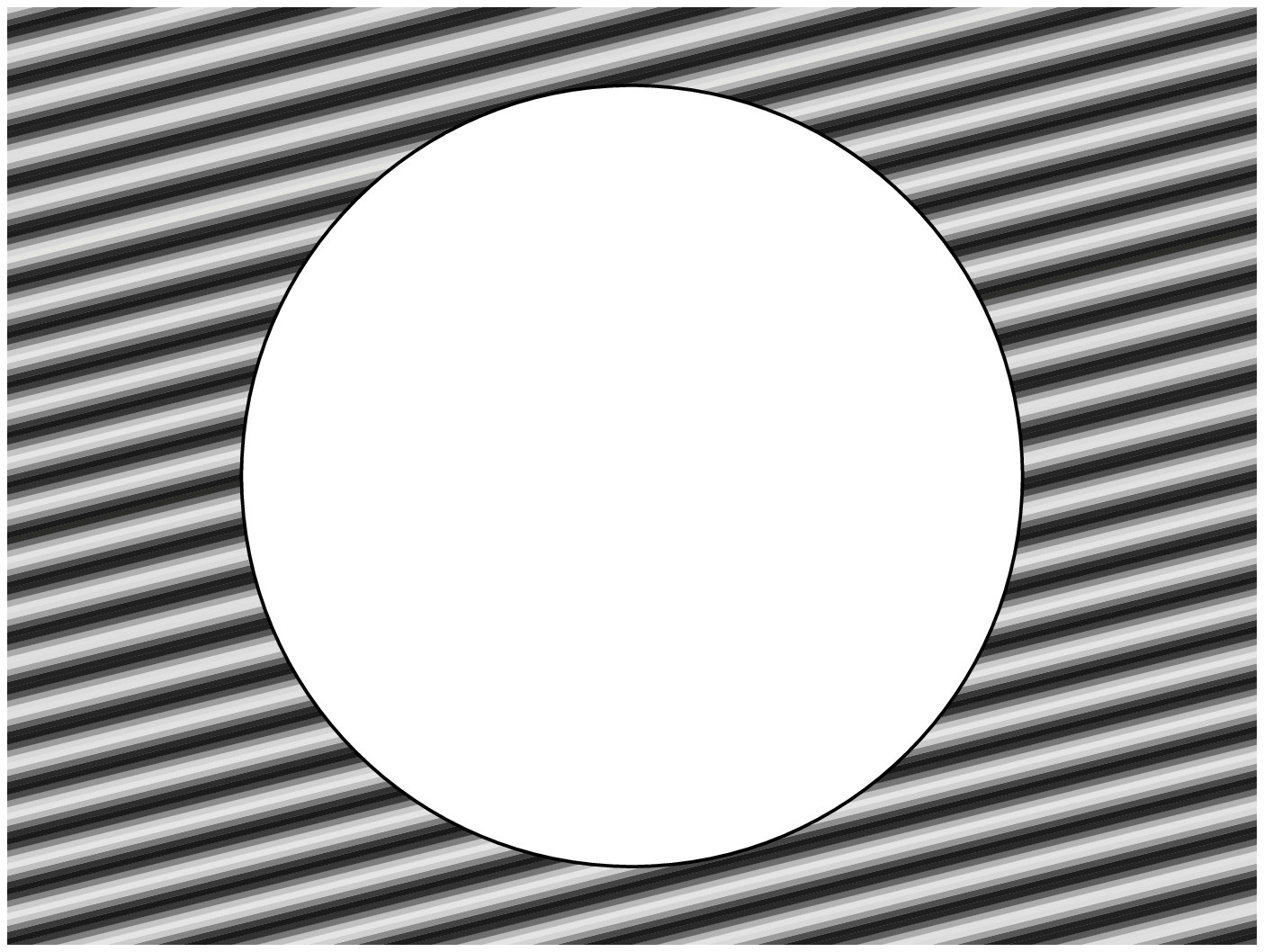}
\vskip 0pt \caption{
{\it Left:} A plane wave on the last-scattering hypersurface.
{\it Right:} The spherical intersection with our past light cone is
shown.\label{fig:plane_wave}}
\end{figure*}

Sachs \& Wolfe (1967) show that a gravitational potential
perturbation produces an anisotropy of the CMB with magnitude
\be
\frac{\Delta T}{T} = \frac{1}{3}\;\frac{\Delta\phi}{c^2},
\ee
where $\Delta\phi$ is evaluated at the intersection of the line-of-sight
and the surface of last scattering (or recombination at $z \approx 1100$).
The $(1/3)$ factor arises because clocks run faster by a factor $(1+\phi/c^2)$
in a gravitational potential, and we can consider the expansion of the
Universe to be a clock.  Since the scale factor is varying as $a
\propto t^{2/3}$ at recombination, the faster expansion leads to a decreased
temperature by $\Delta T/T = -(2/3)\Delta\phi/c^2$, which, when added to the
normal gravitational redshift $\Delta T/T = \Delta\phi/c^2$ yields the $(1/3)$
factor above.
This is an illustration of the ``gauge'' problem in calculating
perturbations in general relativity.
The expected variation of the density contrast as the square of the
scale factor for scales larger than the horizon in the radiation-dominated 
epoch is only obtained after allowance is made for the
effect of the potential on the time.
For a plane wave with wavenumber $k$ we have
$-k^2 \Delta\phi = 4\pi G \Delta\rho$, or
\be
\frac{\Delta\phi}{c^2} = -\frac{3}{2} (H/ck)^2 \frac{\Delta\rho}{\rho_{crit}},
\ee
so when $\rho \approx \rho_{crit}$ at recombination, the Sachs-Wolfe
effect exceeds the physical temperature fluctuation
$\Delta T/T = (1/4)\Delta\rho/\rho$ by a factor
of $2(H/ck)^2$ if fluctuations are adiabatic (all component number densities
varying by the same factor).

In addition to the physical temperature fluctuation and the gravitational
potential fluctuation, there is a Doppler shift term.  When the baryon
fluid has a density contrast given by 
\be
\delta_b(x,t) =
\frac{\Delta\rho_b}{\rho_b} =
\delta_b\exp[ik(x-c_st)],
\ee
where $c_s$ is the sound speed, then
\be
\frac{\partial\Delta\rho_{b}}{\partial t} =
-ikc_s \delta_{b} \rho_{b}
= -\rho_{b} \vec{\nabla}\cdot\vec{\upsilon} =
ik \upsilon \rho_{b}.
\ee
As a result the velocity perturbation is given by
$\upsilon = -c_s\delta_{b}$.
But the sound speed is given by $c_s = \sqrt{\partial P/\partial\rho}
= \sqrt{(4/3)\rho_\gamma c^2/(3\rho_b+4\rho_\gamma)} \approx c/\sqrt{3}$,
since $\rho_\gamma > \rho_b$ at recombination ($z = 1100$).
But the photon density is only slightly higher than the baryon density
at recombination so the sound speed is about 20\% smaller than
$c/\sqrt{3}$.
The Doppler shift term in the anisotropy is given by $\Delta T/T = \upsilon/c$, 
as expected.  This results in $\Delta T/T$ slightly less than
$\delta_b/\sqrt{3}$, which is nearly
$\sqrt{3}$ larger than the physical temperature fluctuation given
by $\Delta T/T = \delta_b/3$.

\begin{figure*}[t]
\includegraphics[width=0.48\columnwidth,angle=0,clip]{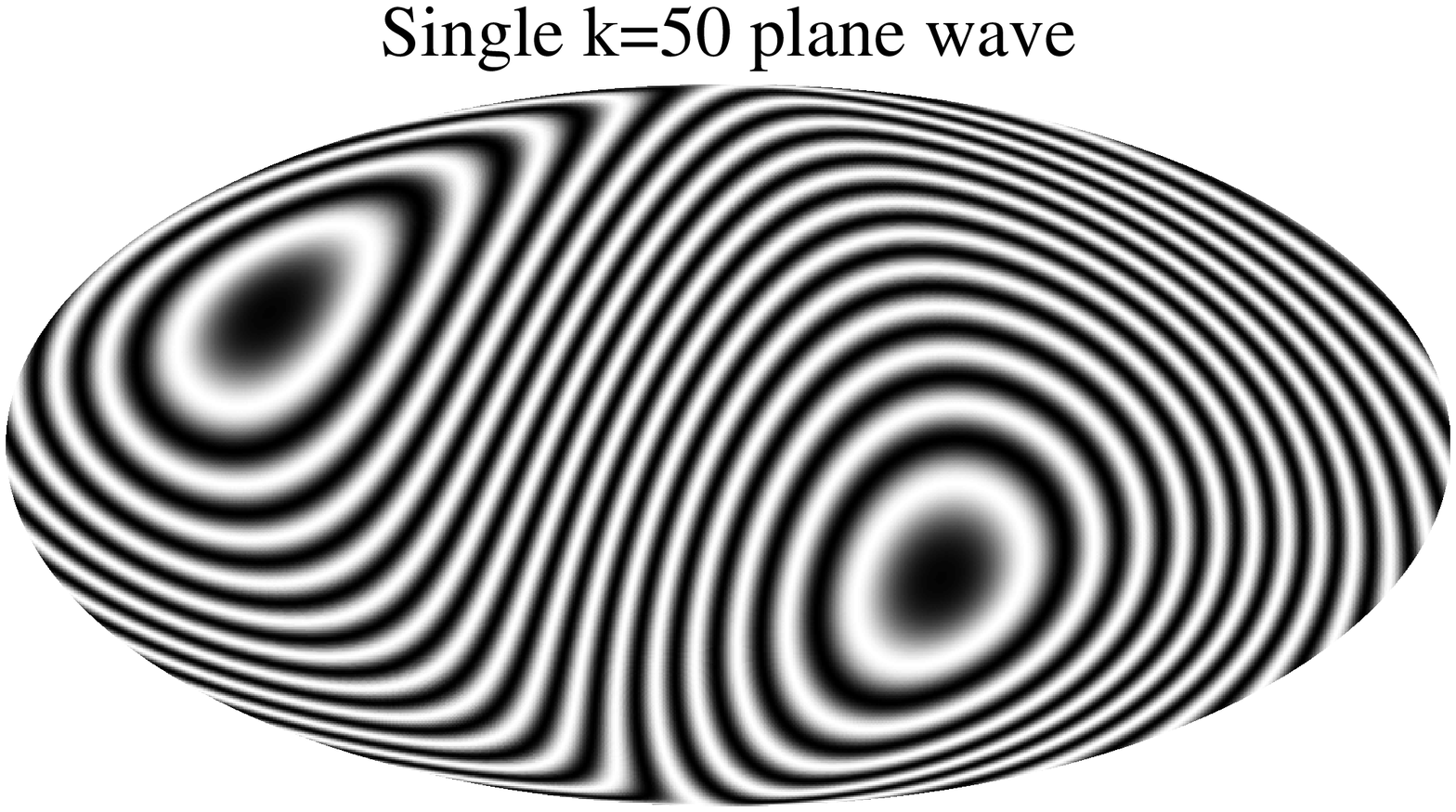} \hfil
\includegraphics[width=0.48\columnwidth,angle=0,clip]{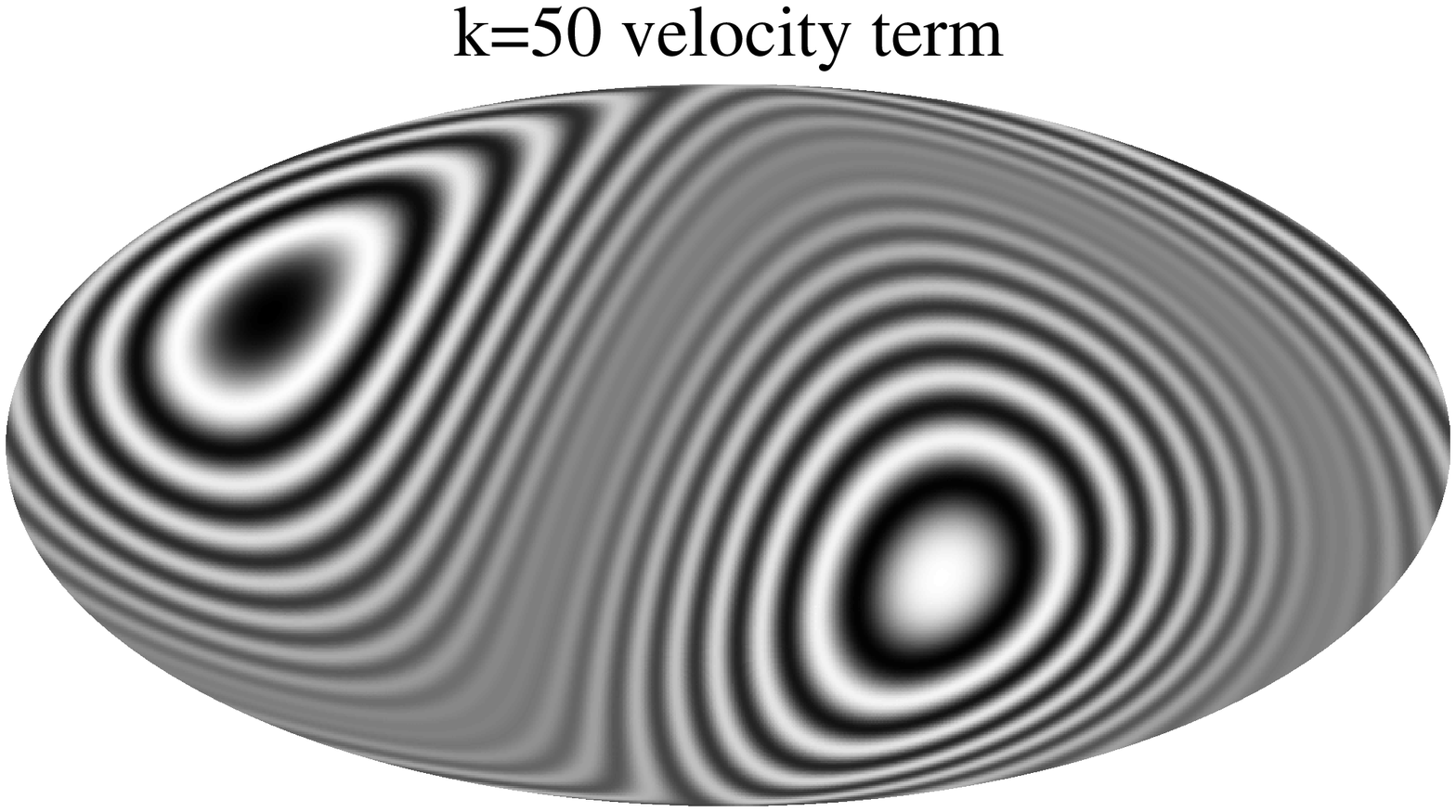} 
\vskip 0pt \caption{
{\it Left:} The scalar density and potential perturbation.
{\it Right:} The vector velocity perturbation.
\label{fig:k_50}}
\end{figure*}

These plane-wave calculations need to be projected onto the sphere that is
the intersection of our past light cone and the hypersurface corresponding
to the time of recombination.  Figure \ref{fig:plane_wave} shows a plane
wave on these surfaces.  The scalar density and potential perturbations
produce a different pattern on the observed sky than the vector velocity
perturbation.  Figure \ref{fig:k_50} shows these patterns on the sky
for a plane wave with $k R_{LS} = 50$, where $R_{LS}$ is the radius of
the last-scattering surface.  The contribution of the velocity term is
multiplied by $\cos\theta$, and since the RMS of this over the
sphere is $\sqrt{1/3}$, the RMS contribution of the velocity term
almost equals the RMS contribution from the density term
since the speed of sound is almost $c/\sqrt{3}$.

The anisotropy is usually expanded in spherical harmonics:
\be
\frac{\Delta T(\hat{n})}{T_0} = \sum_\ell \sum_{m=-\ell}^\ell a_{\ell m}
Y_{\ell m}(\hat{n}).
\ee
Because the Universe is approximately isotropic the probability densities
for all the different $m$'s at a given $\ell$ are identical.  Furthermore, 
the expected value of $\Delta T(\hat{n})$ is obviously zero, and thus the
expected values of the $a_{\ell m}$'s is zero.
But the variance of the $a_{\ell m}$'s is a measurable function of $\ell$, 
defined as
\be
C_\ell = \langle |a_{\ell m}|^2\rangle.
\ee
Note that in this normalization $C_\ell$ and $a_{\ell m}$ are dimensionless.
The harmonic index $\ell$ associated with an angular scale $\theta$ is
given by $\ell \approx 180^\circ/\theta$, but the total number of spherical
harmonics contributing to the anisotropy power at angular scale $\theta$
is given by $\Delta \ell \approx \ell$ times $2\ell + 1$.  Thus to have
equal power on all scales one needs to have 
approximately $C_\ell \propto \ell^{-2}$.  Given that the square of the angular
momentum operator is actually $\ell(\ell+1)$, it is not surprising that
the actual angular power spectrum of the CMB predicted by ``equal power on all
scales'' is
\be
C_\ell = \frac{4\pi \langle Q^2 \rangle}{5 T_0^2}\;\frac{6}{\ell(\ell+1)}, 
\label{eq:hz-cell}
\ee
where $\langle Q^2 \rangle$ or $Q_{rms-PS}^2$ is the expected variance of the
$\ell =2$ component of the sky, which must be divided by $T_0^2$ because
the $a_{\ell m}$'s are defined to be dimensionless.  The ``$4\pi$'' term
arises because the mean of $|Y_{\ell m}|^2$ is $1/(4\pi)$, so the 
$|a_{\ell m}|^2$'s must be $4\pi$ times larger to compensate.  Finally, the
quadrupole has 5 components, while $C_\ell$ is the variance of a single
component, giving the ``5'' in the denominator.  The {\it COBE} DMR experiment
determined $\sqrt{\langle Q^2 \rangle} = 18\;\mu\mbox{K}$, and that the
$C_\ell$'s from $\ell = 2$ to $\ell = 20$ were consistent with
Equation \ref{eq:hz-cell}.

The other common way of describing the anisotropy is in terms of
\be
\Delta T_\ell^2 = \frac{T_0^2\ell(\ell+1)C_\ell}{2\pi}.
\ee
Note these definitions give $\Delta T_2^2 = 2.4 \langle Q^2 \rangle$.
Therefore, the {\it COBE} normalized Harrison-Zel'dovich spectrum has
$\Delta T_\ell^2 = 2.4 \times 18^2 = 778\;\mbox{$\mu$K$^2$}$
for $\ell \leq 20$.

\begin{figure*}[t]
\includegraphics[width=0.48\columnwidth,angle=0,clip]{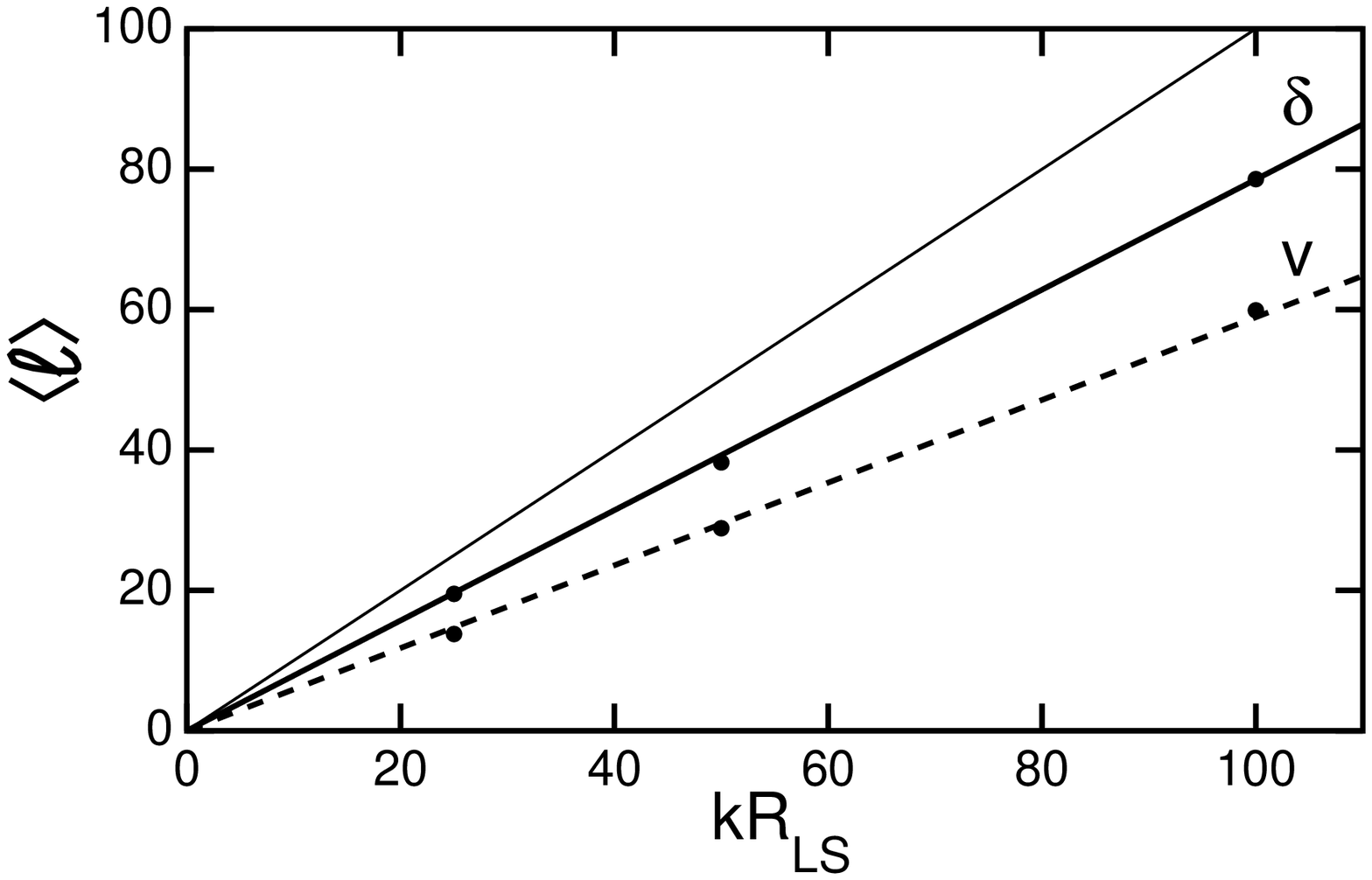} \hfil
\includegraphics[width=0.48\columnwidth,angle=0,clip]{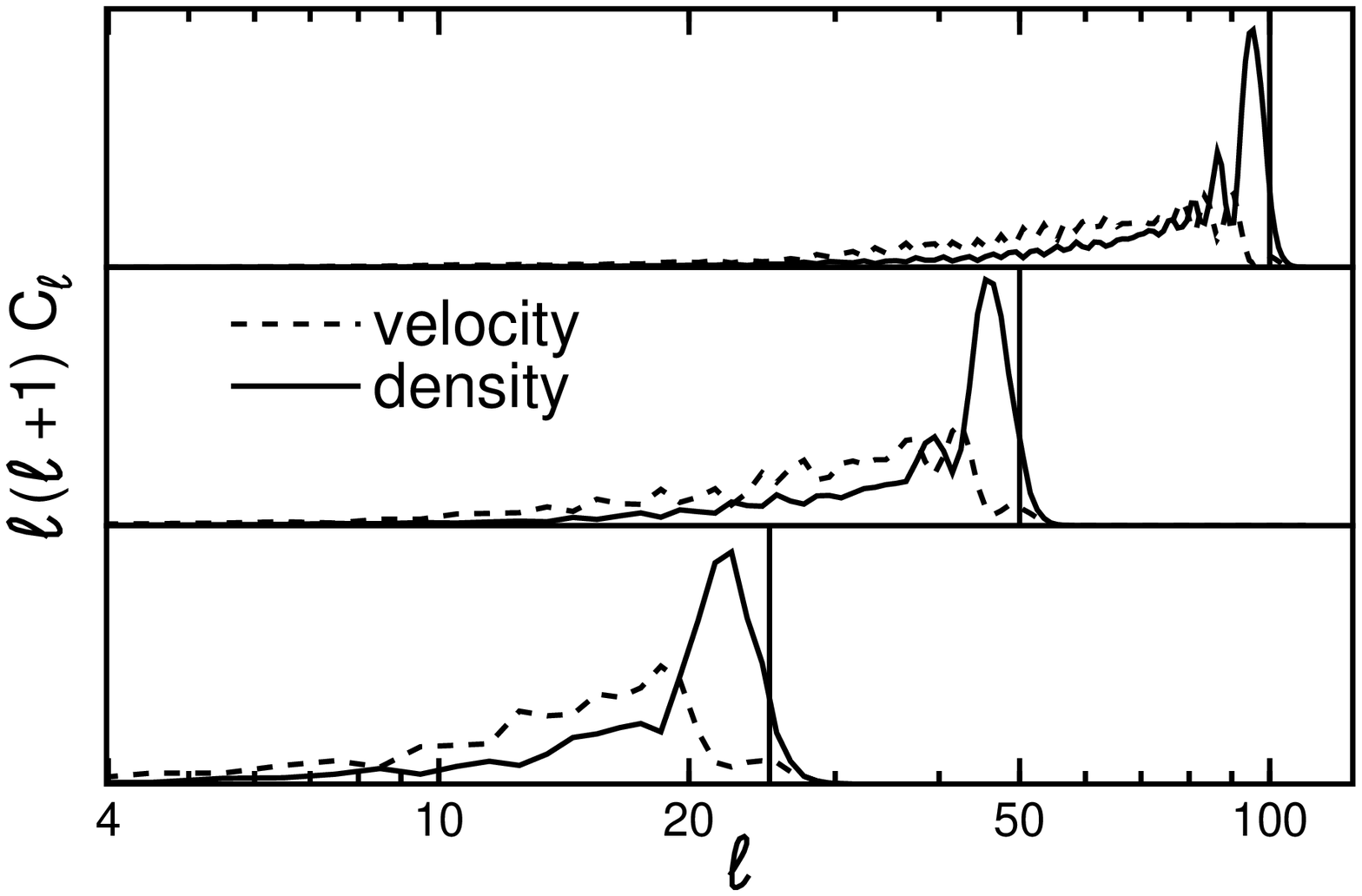}
\vskip 0pt \caption{
{\it Left:} Mean $\ell$ is plotted \vs\ the wavenumber
$kR_{LS}$.  The light solid line shows $\ell = kR_{LS}$, while the
solid line shows $\ell = (\pi/4)kR_{LS}$, and the dashed line
shows $\ell = (3\pi/16)kR_{LS}$.  {\it Right:} The angular power spectra
for single $k$ skies is plotted for $kR_{LS} = 25$, 50, and 100 (from bottom
to top).
\label{fig:ell-vs-k}}
\end{figure*}

It is important to realize that the relationship between the wavenumber
$k$ and the spherical harmonic index $\ell$ is not a simple
$\ell = k R_{LS}$.  Figure \ref{fig:k_50} shows that while
$\ell = k R_{LS}$ at the ``equator'' the poles have lower $\ell$.
In fact, if $\mu = \cos\theta$, where $\theta$ is the angle between
the wave vector and the line-of-sight, then the ``local $\ell$'' is
given by $k R_{LS} \sqrt{1-\mu^2}$.  The average of this over
the sphere is $\langle \ell \rangle = (\pi/4) k R_{LS}$.
For the velocity term the power goes to zero when $\mu = 0$ on
the equator, so the average $\ell$ is smaller,
$\langle \ell \rangle = (3\pi/16) k R_{LS}$, and the distribution of
power over $\ell$ lacks the sharp cusp at $\ell = k R_{LS}$.
As a result the velocity term, while contributing about 60\% as much to the
RMS anisotropy as the density term, does not contribute this
much to the peak structure in the angular power spectrum.
Thus the old nomenclature of ``Doppler'' peaks was not
appropriate, and the new usage of ``acoustic'' peaks is more
correct.  Figure \ref{fig:ell-vs-k} shows the angular power spectrum
from single $k$ skies for both the density and velocity terms for
several values of $k$, and a graph of the variance-weighted mean
$\ell$ \vs\ $kR_{LS}$.  These curves were computed numerically but
have the expected forms given by the spherical Bessel function
$j_\ell$ for the density term and $j_\ell^\prime$ for the velocity
term.

\begin{figure*}[t]
\includegraphics[width=0.48\columnwidth,angle=0,clip]{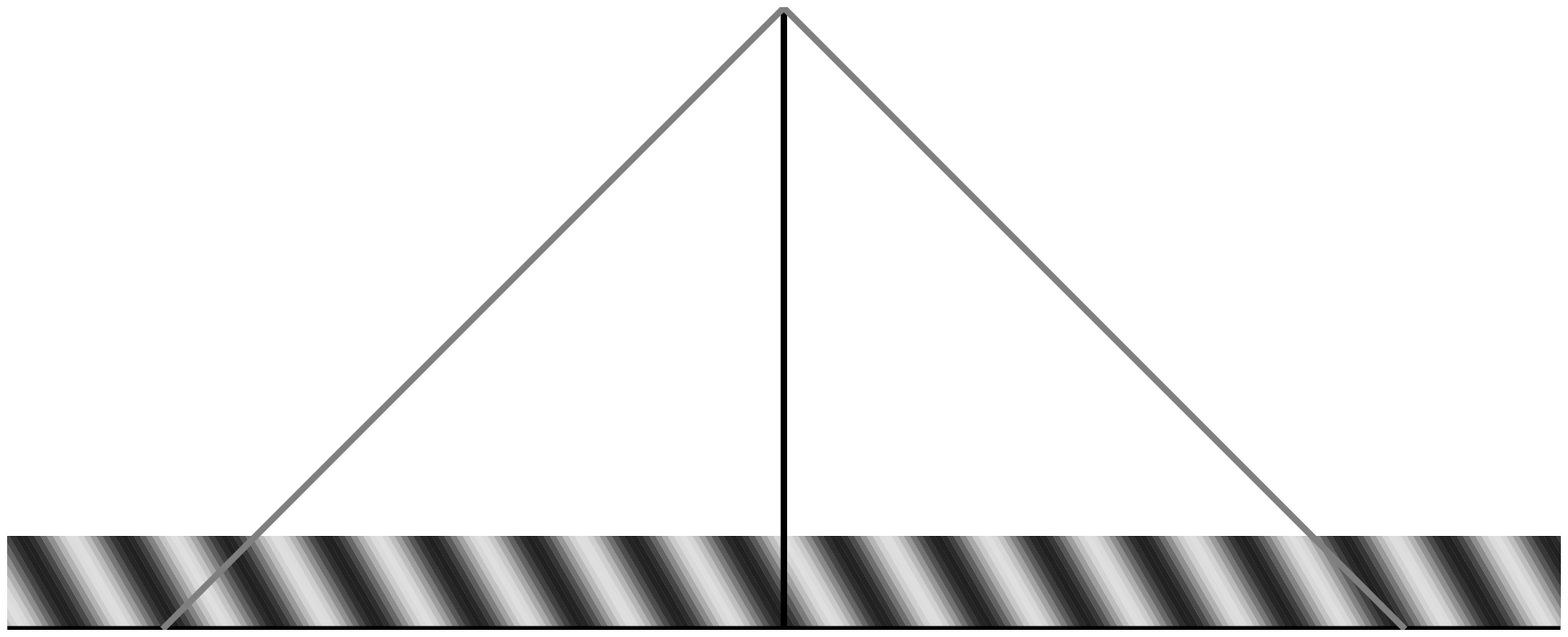} \hfil
\includegraphics[width=0.48\columnwidth,angle=0,clip]{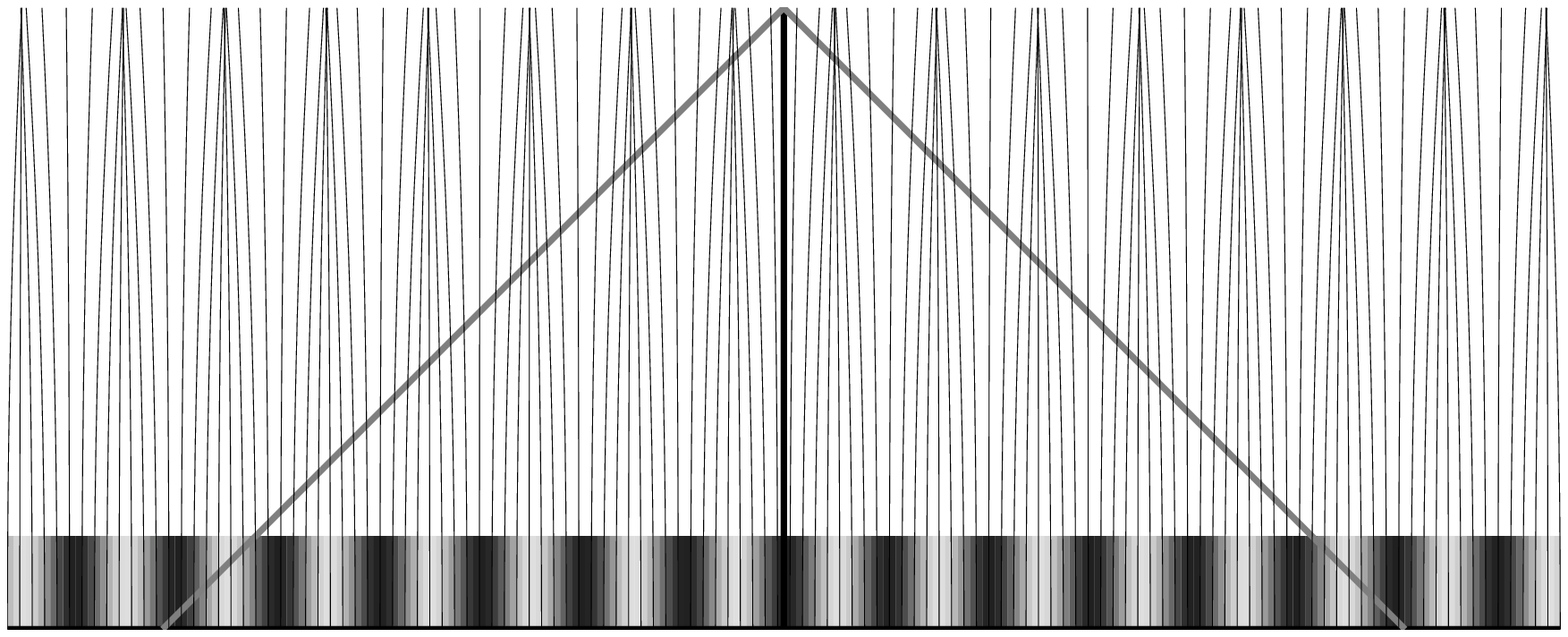}
\vskip 0pt \caption{
On the left a conformal spacetime diagram showing
a traveling wave in the baryon-photon
fluid.  On the right, the stationary CDM wave and the world
lines of matter falling into the potential wells.  For this
wavenumber the density contrast in the baryon-photon
fluid has undergone one-half cycle of its oscillation and
is thus in phase with the Sachs-Wolfe effect from the CDM.
This condition defines the first acoustic peak.\label{fig:waves}}
\end{figure*}

Seljak (1994) considered a simple model in which the photons
and baryons are locked together before recombination, and
completely noninteracting after recombination.  Thus the
opacity went from infinity to zero instantaneously.  Prior
to recombination there were two fluids, the photon-baryon
fluid and the CDM fluid, which interacted only 
gravitationally.  The baryon-photon fluid has a sound speed
of about $c/\sqrt{3}$ while the dark matter fluid has a
sound speed of zero.  Figure \ref{fig:waves}
shows a conformal spacetime diagrams with a traveling
wave in the baryon-photon fluid and the stationary wave
in the CDM.  The CDM dominates the potential, so
the large-scale structure (LSS) forms in the potential wells
defined by the CDM.

\begin{figure*}[b!]
\includegraphics[width=1.00\columnwidth,angle=0,clip]{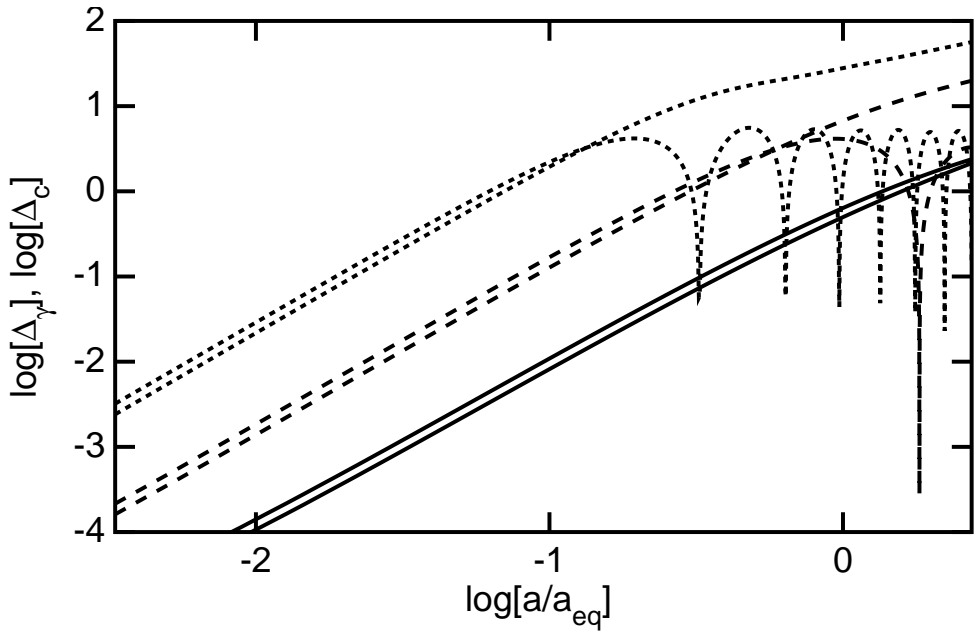} 
\vskip 0pt \caption{
Density contrasts in the CDM and the photons for wavenumbers $\kappa = 5$, 20, 
and 80 (see Fig. \ref{fig:delta-vs-k}) as a function of the scale factor 
relative to the scale factor and when matter and radiation densities
were equal.  The photon density contrast starts out slightly larger
than the CDM density contrast but oscillates.
\label{fig:delta-vs-a}}
\end{figure*}

In Seljak's simple two-fluid model, there are five variables to
follow: the density contrast in the CDM and baryons,
$\delta_c$ and $\delta_b$, the velocities of these fluids $\upsilon_c$
and $\upsilon_b$, and the potential $\phi$.  The photon density contrast is
$(4/3)\delta_b$.  In Figure \ref{fig:delta-vs-a}
the density contrasts are plotted \vs\ the scale factor for several
values of the wavenumber.  To make this plot the density contrasts
were adjusted for the effect of the potential on the time, with
\be
\Delta_c = \delta_c + 3 H \int (\phi/c^2) dt
\ee
and
\be
\Delta_\gamma = \delta_\gamma + 4 H \int (\phi/c^2) dt.
\ee
Remembering that $\phi$ is negative when $\delta$ is positive, the
two terms on the right-hand side of the above equations cancel almost entirely
at early times, leaving a small residual growing like $a^2$ prior to
$a_{eq}$, the scale factor when the matter density and the radiation
density were equal.  Thus these adjusted density contrasts evolve like
$\Omega^{-1}-1$ in homogeneous universes.

\begin{figure*}[t]
\includegraphics[width=1.00\columnwidth,angle=0,clip]{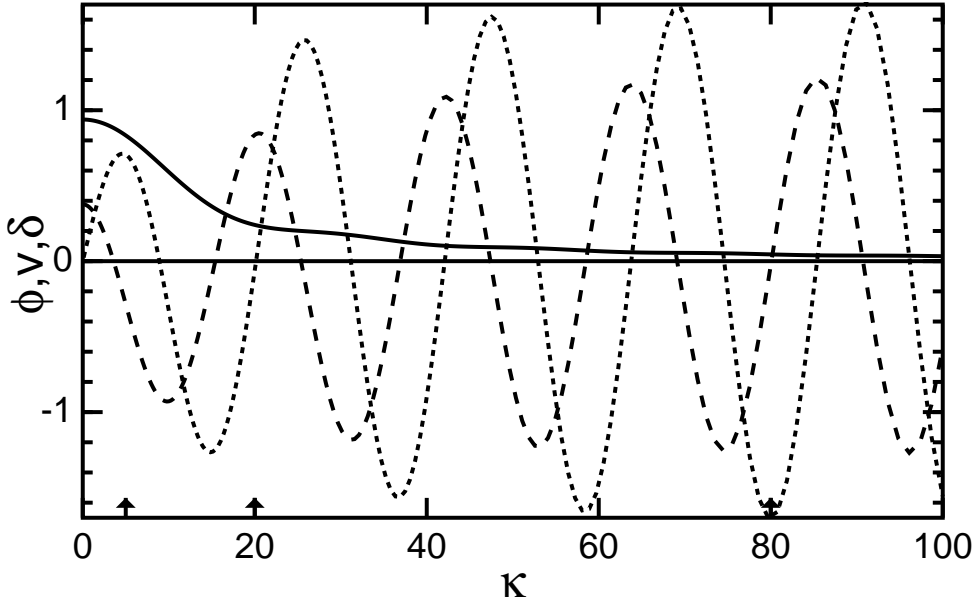}
\vskip 0pt \caption{
Density contrasts at recombination as a function of
wavenumber $\kappa$.  The arrows on the x-axis indicate the values
of $\kappa$ for $\delta$ \vs\ $a$, as plotted in Figure
\protect\ref{fig:delta-vs-a}.  The solid curve shows the potential
(the initial potential is always $\phi = 1$), the long dashed curve
curve shows the combined potential plus density effect on the
CMB temperature, while the short dashed curve shows the velocity
of the baryon-photon fluid.\label{fig:delta-vs-k}}
\end{figure*}

Figure \ref{fig:delta-vs-k} shows the potential that survives to
recombination and produces LSS, the potential plus
density effect on the CMB temperature, and the velocity of the baryons
as function of wavenumber.  Close scrutiny of the potential curve
in the plot shows the baryonic wiggles in the LSS
that may be detectable in the large redshift surveys by the 2dF and SDSS
groups.

\begin{figure*}[t]
\includegraphics[width=1.00\columnwidth,angle=0,clip]{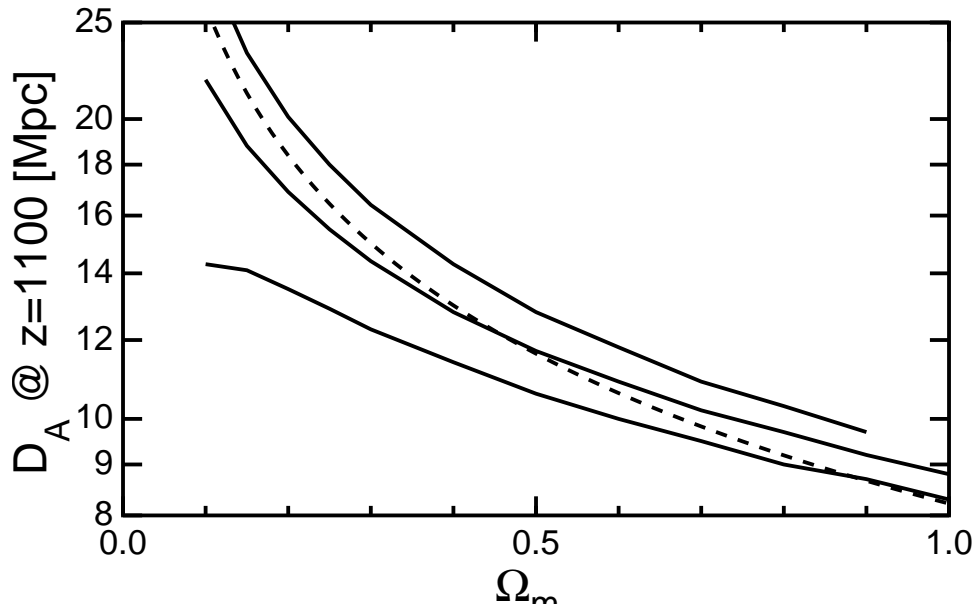}
\vskip 0pt \caption{
The angular size distance \vs\ $\Omega_m$ for $H_0 = 60$ \kmsmpc\
and three different values of $\Omega_{tot} (0.9$, 1, and 1.1, from top
to bottom).  The dashed curve shows $\Omega_m^{-1/2}$.\label{fig:D_A}}
\end{figure*}

A careful examination of the angular power spectrum allows several
cosmological parameters to be derived.  The baryon to photon ratio
and the dark matter to baryon density ratio can both be derived from
the amplitudes of the first two acoustic peaks.  Since the photon
density is known precisely, the peak amplitudes determine
the baryon density $\omega_b = \Omega_b h^2$ and the cold dark matter
density $\omega_c = \Omega_{CDM} h^2$.
The matter density is given by $\Omega_m = \Omega_b + \Omega_{CDM}$.
The amplitude $\langle Q^2 \rangle$
and spectral index $n$ of the primordial density perturbations
are also easily observed.
Finally the angular scale of the peaks depends on the ratio of the
angular size distance at recombination to the distance sound can travel
before recombination.  Since the speed of sound is close to
$c/\sqrt{3}$, this sound travel distance is primarily affected by the age
of the Universe at $z = 1100$.  The age of the Universe goes like
$t \propto \rho^{-1/2} \propto \Omega_m^{-1/2} h^{-1}$.
The angular size distance is proportional to $h^{-1}$ as well, so
the Hubble constant cancels out.  The angular size distance is
almost proportional to $\Omega_m^{-1/2}$, but this relation is
not quite exact.  Figure \ref{fig:D_A} compares the angular size distance
to $\Omega_m^{-1/2}$.  One sees that a peak position that
corresponds to $\Omega_{tot} = 0.95$ if $\Omega_m = 0.2$ can also
be fit by $\Omega_{tot} = 1.1$ if $\Omega_m = 1$.  Thus, to first
order the peak position is a good measure of $\Omega_{tot}$.

The CMBFAST code by Seljak \& Zaldarriaga (1996) provides the ability
to quickly compute the angular power spectrum $C_\ell$.  Typically
CMBFAST runs in about 1 minute for a given set of cosmological
parameters.  However, two different groups have developed even
faster methods to evaluate $C_\ell$.   Kaplinghat, Knox, \& Skordis
(2002) have published the Davis Anisotropy Shortcut (DASh), with
code available for download.  This program interpolates among
precomputed $C_\ell$'s.
Kosowsky, Milosavljevi\'c, \& Jiminez (2002) discuss 
combinations of the parameters that produce simple changes
in the power spectrum, and also allow accurate and fast
interpolation between $C_\ell$'s.  These shortcuts allow the
computation of a $C_\ell$ from model parameters in about
1 second.  This allows the rapid computation of the
likelihood of a given data set $D$ for a set of model
parameters $M$, $L(D|M)$.  When computing the likelihood
for high signal-to-noise ratio observations of a small area of the sky, biases
due to the non-Gaussian shape of the likelihood are common.
This can be avoided using the offset log-normal form for the
likelihood $L(C_\ell)$ advocated by Bond, Jaffe, \& Knox (2000).

\begin{table}
\caption{Beam Size and Calibration Corrections
\label{tab:cal}}
\begin{tabular}{lccc}
\hline \hline
Experiment & $\theta_B$ (${}^\prime$) &
$100(\Delta\theta)/\theta$ & $100(\Delta dT)/dT$ \\
\hline
COBE      & 420.0 & $-$0.3 &   ...   \\
ARCHEOPS  &  15.0 & $-$0.2 &     2.7 \\
BOOMERanG &  12.9 &   10.5 &  $-$7.6 \\
MAXIMA    &  10.0 & $-$0.9 &     0.2 \\
DASI      &   5.0 & $-$0.4 &     0.7 \\
VSA       &   3.0 & $-$0.5 &  $-$1.7 \\
CBI       &   1.5 & $-$1.1 &  $-$0.2 \\
\hline
\end{tabular}
\end{table}

The likelihood is a probability distribution over the data,
so $\int L(D|M) dD = 1$ for any $M$.  It is not a 
probability distribution over the models, so one should
never attempt to evaluate $\int L dM$.  
For example, one could consider the likelihood as a function
of the model parameters $H_0$ in \kmsmpc\ and
$\Omega_m$ for flat $\Lambda$CDM models, or one
could use the parameters $t_0$ in seconds and
$\Omega_m$.  For any $(H_0,\Omega_m)$ there is a
corresponding $(t_0,\Omega_m)$ that makes exactly
the same predictions, and therefore gives the same
likelihood.  But the integral of the likelihood over 
$dt_0 d\Omega_m$ will be much larger than the integral
of the likelihood over $dH_0 d\Omega_m$ just because
of the Jacobian of the transformation between the different
parameter sets.

Wright (1994) gave the example of determining the 
primordial power spectrum power-law index $n$,
$P(k) = A(k/k_0)^n$.  Marginalizing over the
amplitude by integrating the likelihood over $A$
gives very different results for different values of
$k_0$.  Thus, it is very unfortunate that
Hu \& Dodelson (2002) still accept integration
over the likelihood.

Instead of integrating over the likelihood one
needs to define the {\it a posteriori} probability of the
models $p_f(M)$ based on an {\it a priori} distribution 
$p_i(M)$ and Bayes' theorem:
\be
p_f(M) \propto p_i(M)L(D|M).
\ee
It is allowable to integrate $p_f$ over the space of models
because the prior will transform when changing variables
so as to keep the integral invariant.

In the modeling reported here, the {\it a priori} distribution is chosen
to be uniform in $\omega_b$, $\omega_c$, $n$, $\Omega_V$, $\Omega_{tot}$,
and $z_{ri}$.  
In doing the fits, the model $C_\ell$'s are adjusted by a
factor of $\exp[a+b\ell(\ell+1)]$ before comparison with the data.  Here $a$
is a calibration adjustment, and $b$ is a beam size correction that assumes
a Gaussian beam.
For {\it COBE}, $a$ is the overall amplitude scaling parameter instead of
a calibration correction.
Marginalization over the calibration and beam size corrections 
for each experiment, and the overall spectral amplitude, is done by
maximizing the likelihood, not by integrating the likelihood.  
Table \ref{tab:cal} gives these beam and calibration corrections for
each experiments.  All of these corrections are less than the quoted
uncertainties for these experiments.
BOOMERanG stands out in the table for having honestly reported its 
uncertainties: $\pm 11\%$ for the beam size and $\pm 10\%$ for the gain.
The likelihood is given by
\bea
-2 \ln L = \chi^2 & = & \sum_j \Bigl\{ f(a_j/\sigma[a_j]) + f(b_j/\sigma[b_j]) 
\nonumber\\
& + &  \sum_i f([Z_{ij}^o-Z_{ij}^c]/\sigma[Z_{ij}]) \Bigr\},
\eea
where $j$ indexes over experiments, $i$ indexes over points within
each experiment, $Z = \ln(C_\ell + N_\ell)$ in the offset log normal
approach of Bond \etal\ (2000), and $N_\ell$ is the noise bias.
Since for {\it COBE} $a$ is the overall normalization, $\sigma(a)$ is set
to infinity for this term to eliminate it from the likelihood.
The function $f(x)$ is $x^2$ for small $|x|$ but switches to
$4(|x|-1)$ when $|x| > 2$.  This downweighs outliers in the data.
Most of the experiments have double tabulated their data.  I have used
both the even and odd points in my fits, but I have multiplied the
$\sigma$'s by $\sqrt{2}$ to compensate.  Thus, I expect to get $\chi^2$ per
degree of freedom close to 0.5 but should have the correct sensitivity to
cosmic parameters.

\begin{table}
\caption{Cosmic Parameters from pre-{\it WMAP} CMB Data only
\label{tab:param}}
\begin{tabular}{ccccccccc}
\hline \hline
Parameter &&& Mean &&& $\sigma$ && Units \\
\hline
$\omega_b$ &&&  0.0206  &&& 0.0020 && \\
$\omega_c$ &&&  0.1343  &&& 0.0221 && \\
$\Omega_V$ &&&  0.3947  &&& 0.2083 && \\
$\Omega_k$ &&&$-$0.0506 &&& 0.0560 && \\
$z_{ri}$   &&&     7.58 &&& 3.97   && \\
$n$        &&&   0.9409 &&& 0.0379 && \\
$H_0$  &&&  51.78   &&& 12.26  && \kmsmpc \\
$t_0$  &&&  15.34   &&& 1.60   && Gyr \\
$\Gamma$   &&&  0.2600  &&& 0.0498 && \\
\hline \hline
\end{tabular}
\end{table}

The scientific results such as the mean values and the
covariance matrix of the parameters can be
determined by integrations over parameter space
weighted by $p_f$.  
Table \ref{tab:param} shows the mean and
standard deviation of the parameters determined by
integrating over the {\it a posteriori} probability distribution of
the models.
The evaluation of integrals
over multi-dimensional spaces can require a large
number of function evaluations when the dimensionality
of the model space is large, so a Monte Carlo
approach can be used.  
To achieve an accuracy of ${\cal O}(\epsilon)$ in
a Monte Carlo integration requires ${\cal O}(\epsilon^{-2})$
function evaluations, while achieving the same
accuracy with a gridding approach requires
${\cal O}(\epsilon^{-n/2})$ evaluations when second-order methods are
applied on each axis.  The Monte Carlo approach is more
efficient for more than four dimensions.
When the CMB data get better, the likelihood gets more
and more sharply peaked as a function of the parameters,
so a Gaussian approximation to $L(M)$ becomes more
accurate, and concerns about banana-shaped confidence
intervals and long tails in the likelihood are reduced.
The Monte Carlo Markov Chain  (MCMC)
approach using the Metropolis-Hastings algorithm to
generate models drawn from $p_f$ is a relatively
fast way to evaluate these integrals (Lewis \&
Bridle 2002).  In the MCMC, a ``trial'' set of parameters
is sampled from the proposal density $p_t(P^\prime;P)$,
where $P$ is the current location in parameter space,
and $P^\prime$ is the new location.  Then the trial location is
accepted with a probability given by
\be
\lambda = \frac{p_f(P^\prime)}{p_f(P)} \frac{p_t(P;P^\prime)}{p_t(P^\prime;P)}.
\ee
When a trial is accepted the Markov chain one sets
$P = P^\prime$.  This algorithm guarantees that the accepted
points in parameter space are sampled from the
{\it a posteriori} probability distribution.

\begin{figure*}[t]
\includegraphics[width=0.48\columnwidth,angle=0,clip]{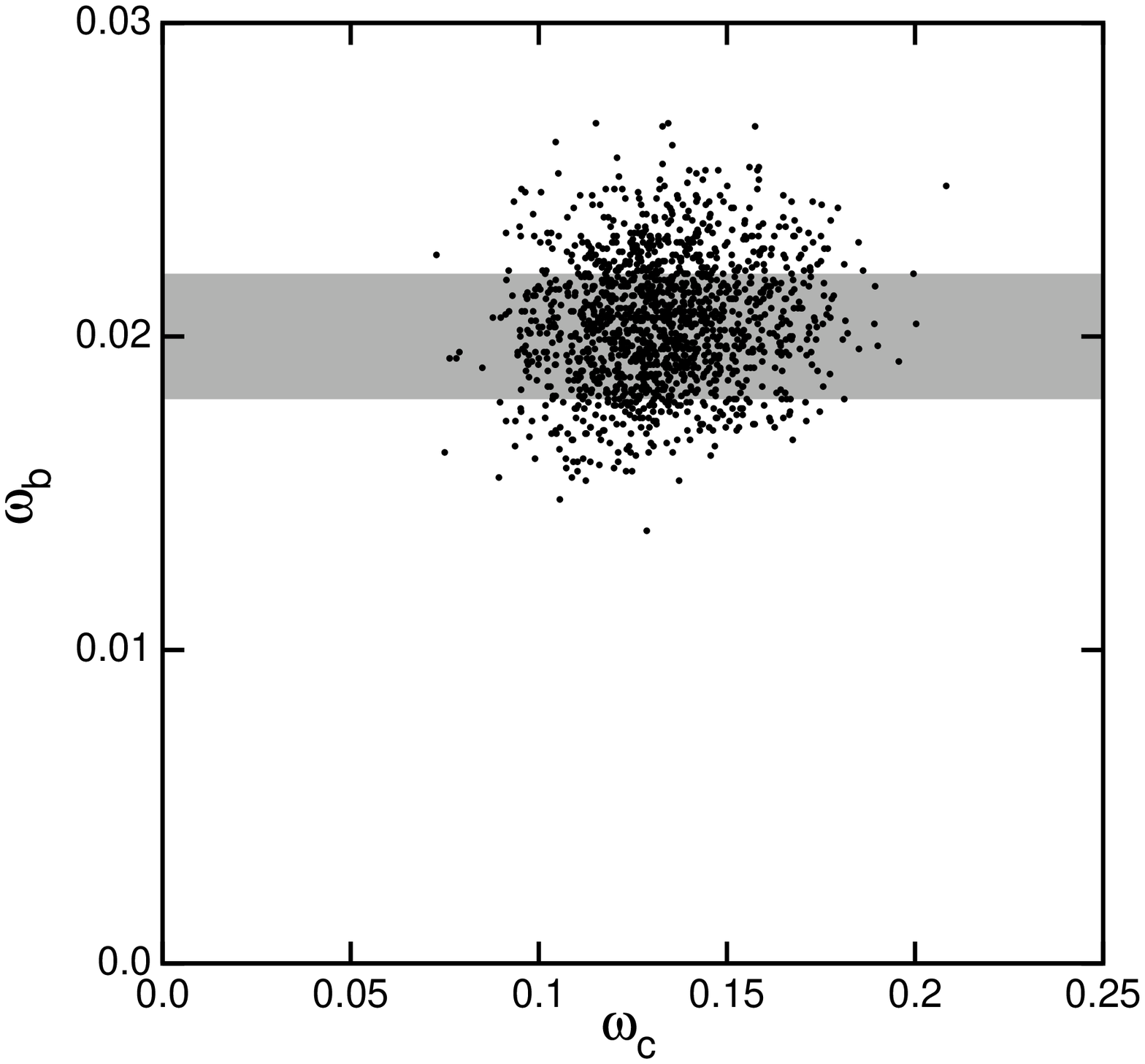} \hfill
\includegraphics[width=0.48\columnwidth,angle=0,clip]{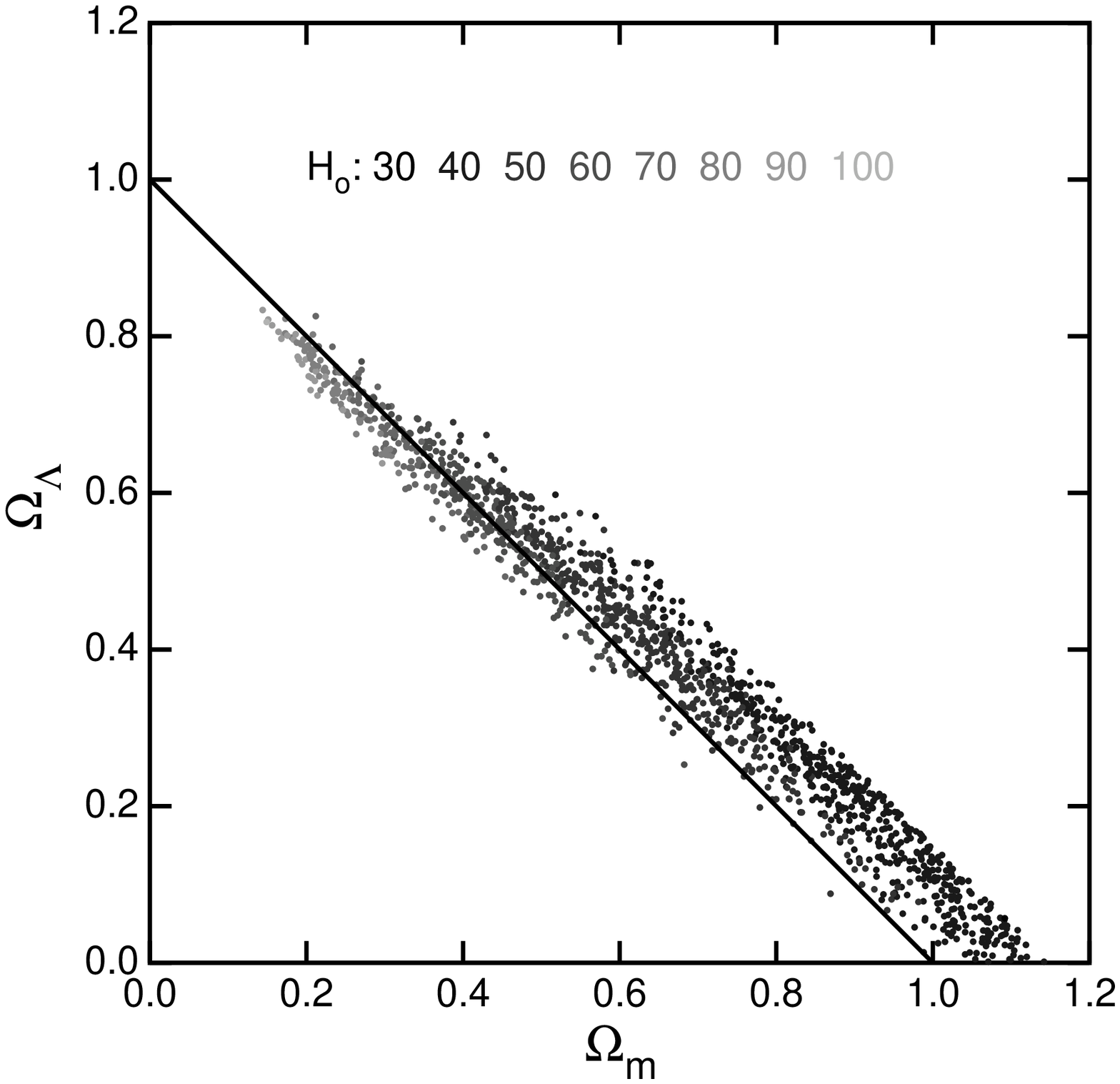}
\vskip 0pt \caption{
{\it Left:} Clouds of models drawn from the {\it a posteriori} distribution
based on the CMB data set as of 19 November 2002.  The gray band shows
the Big Bang nucleosynthesis determination of $\omega_b$ ($\pm2\sigma)$
from Burles, Nollett, \& Turner (2001).  {\it Right:} the same set of models
in the $\Omega_m$, $\Omega_\Lambda$ plane.
\label{fig:wb-wc}}
\end{figure*}

\begin{figure*}[b!]
\includegraphics[width=1.00\columnwidth,angle=0,clip]{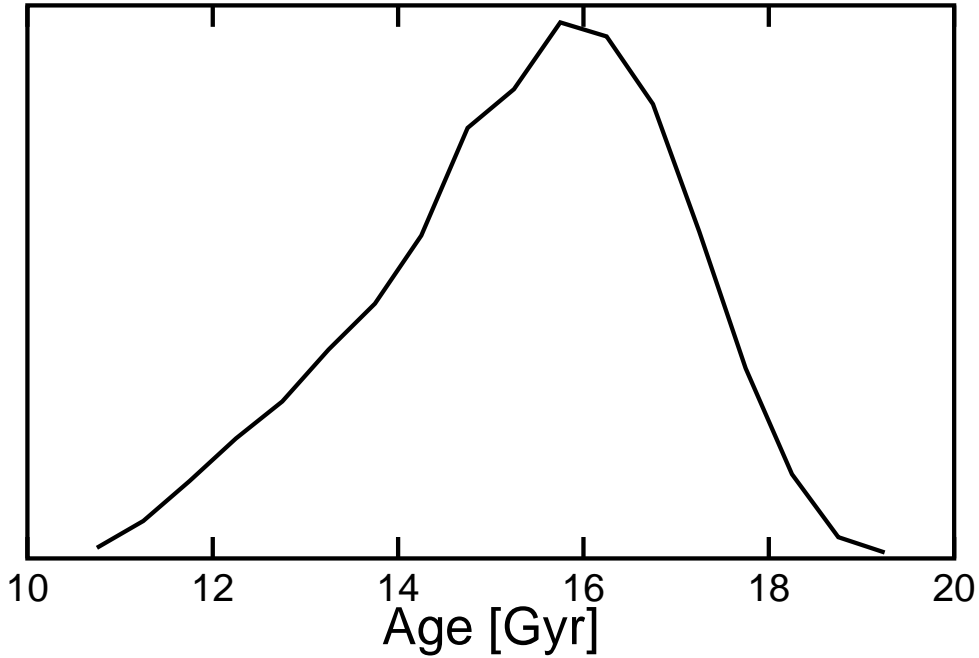}
\vskip 0pt \caption{
Distribution of the age of the Universe based only
on the pre-{\it WMAP} CMB data.
\label{fig:t0dist}}
\end{figure*}

The most common choice for the proposal density is
one that depends only on the parameter change $P^\prime - P$.
If the proposal density is a symmetric function then the
ratio $p_t(P;P^\prime)/p_t{P^\prime;P)}$ = 1 and $\lambda$ is
then just the ratio of {\it a posteriori} probabilities.  But the most
efficient choice for the proposal density is $p_f(P)$ which is not
a function of the parameter change, because 
this choice makes $\lambda =1$ and all trials are accepted.
However, if one knew how to sample models from $p_f$, why
waste time calculating the likelihoods?

Just plotting the cloud of points
from MCMC gives a useful indication of the
allowable parameter ranges that are consistent with
the data.  I have done some MCMC calculations using
the DASh (Kaplinghat \etal\ 2002) to find the $C_\ell$'s.
I found DASh to be user unfriendly and too likely to terminate
instead of reporting an error for out-of-bounds parameter sets,
but it was fast.
Figure \ref{fig:wb-wc} shows the range
of baryon and CDM densities consistent with the 
CMB data set from {\it COBE} (Bennett \etal\ 1996), 
ARCHEOPS (Amblard 2003), 
BOOMERanG (Netterfield \etal\ 2002), MAXIMA (Lee \etal\ 2001),
DASI (Halverson \etal\ 2002), VSA (Scott \etal\ 2003),
and CBI (Pearson \etal\ 2003), 
and the range of matter and vacuum densities consistent with these
data.  The Hubble constant is strongly correlated with
position on this diagram.  Figure \ref{fig:t0dist} shows the
distribution of $t_0$ for models consistent with this
pre-{\it WMAP} CMB data set.  The relative uncertainty in $t_0$
is much smaller than the relative uncertainty in $H_0$
because the low-$H_0$ models have low vacuum energy density ($\Omega_V$),
and thus low values of the product $H_0 t_0$.  The CMB
data are giving a reasonable value for $t_0$ without using
information on the distances or ages of objects, which is
an interesting confirmation of the Big Bang model.

\begin{figure*}[t]
\includegraphics[width=1.00\columnwidth,angle=0,clip]{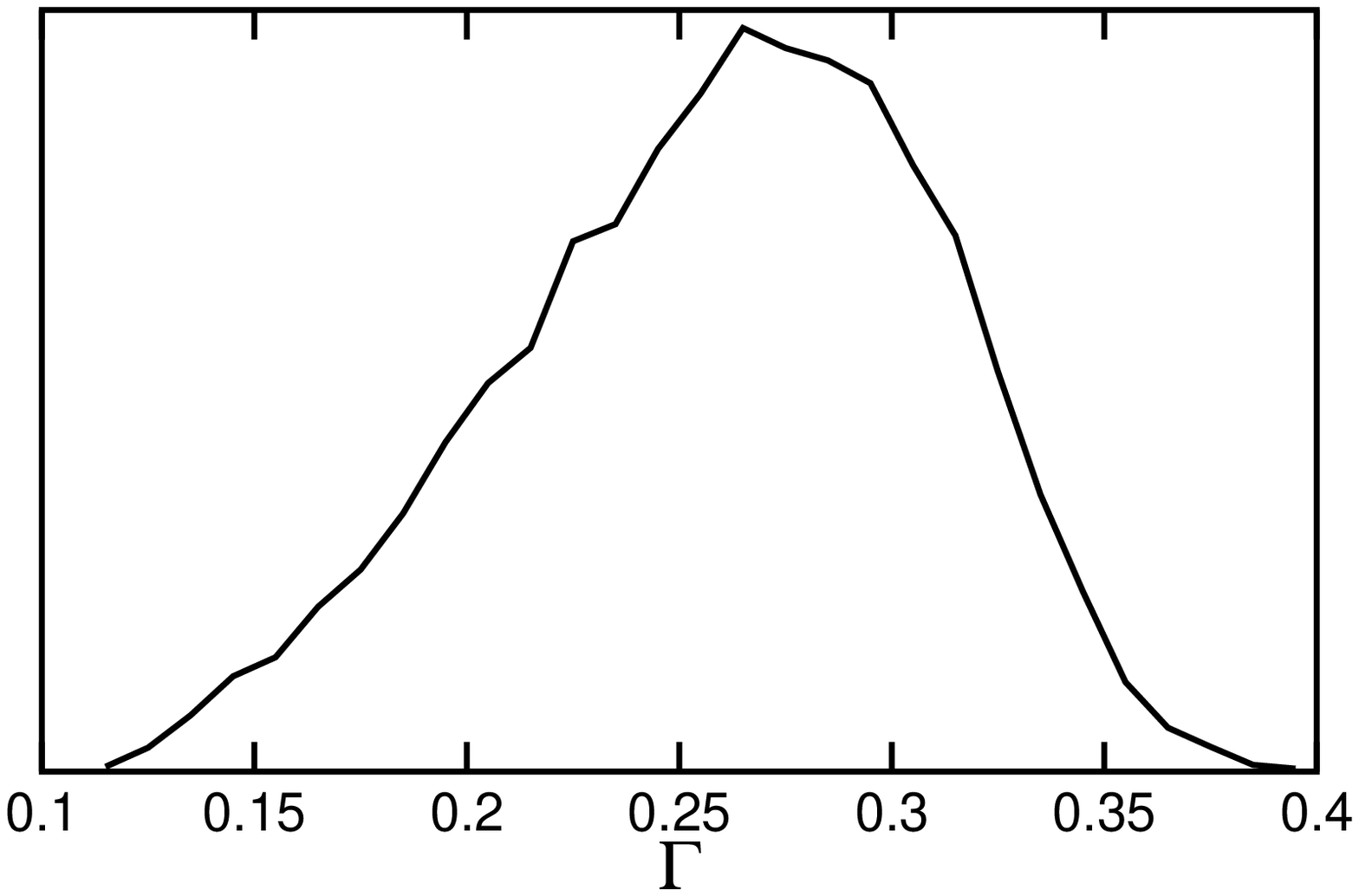}
\vskip 0pt \caption{
Distribution of the LSS power spectrum shape parameter
$\Gamma = \Omega_m h \exp(-2\Omega_b)$ from the pre-{\it WMAP} CMB data.
\label{fig:Gammadist}}
\end{figure*}

\begin{figure*}[t]
\includegraphics[width=1.00\columnwidth,angle=0,clip]{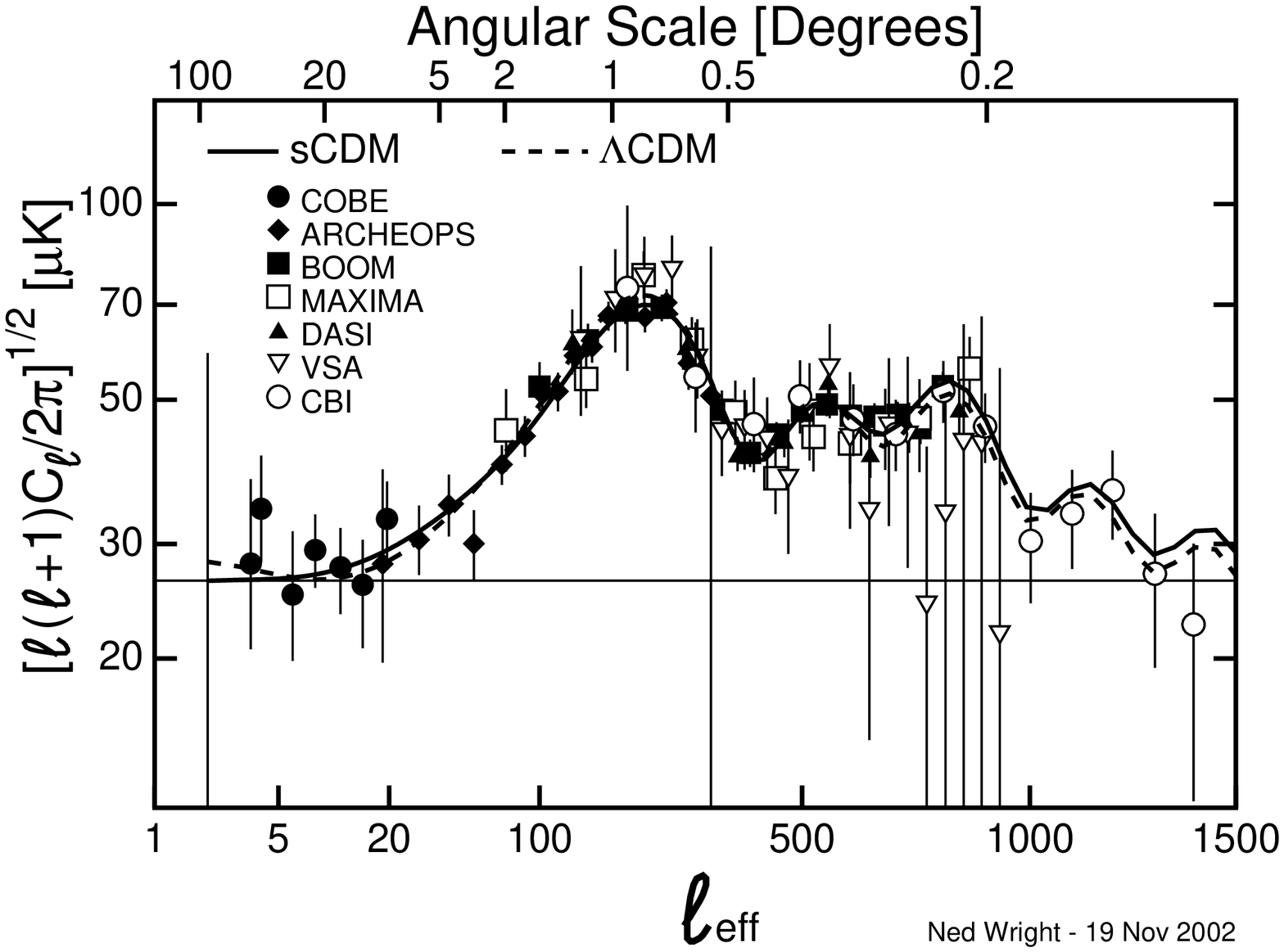}
\vskip 0pt \caption{
Two flat $n=1$ models.  One shows $\Lambda$CDM with $\Omega_\Lambda = 2/3$.
The best fit gives $\omega_b = 0.022$ and $\omega_c = 0.132$, implying
$H_0 = 68$ \kmsmpc.   The other fit shows $\Omega_\Lambda = 0$ with
$\omega_b = 0.021$ and $\omega_c =0.196$, implying $H_0 = 47$ \kmsmpc.
\label{fig:CMB-19Nov02}}
\end{figure*}

Peacock \& Dodds (1994) define a shape parameter for the observed
LSS power spectrum, $\Gamma = \Omega_m h \exp(-2\Omega_b)$.
There are other slightly different definitions of $\Gamma$ in use, but 
this will be used consistently here.
Peacock \& Dodds determine $\Gamma = 0.255 \pm 0.017 + 0.32(n^{-1}-1)$.
The CMB data specify $n$, so the slope correction in the last term is
only $0.020 \pm 0.013$.  Hence, the LSS power spectrum wants
$\Gamma = 0.275 \pm 0.02$.  The models based only on the pre-{\it WMAP}\ CMB
data give the distribution in $\Gamma$ shown in Figure \ref{fig:Gammadist},
which is clearly consistent with the LSS data.

\begin{figure*}[t]
\includegraphics[width=1.00\columnwidth,angle=0,clip]{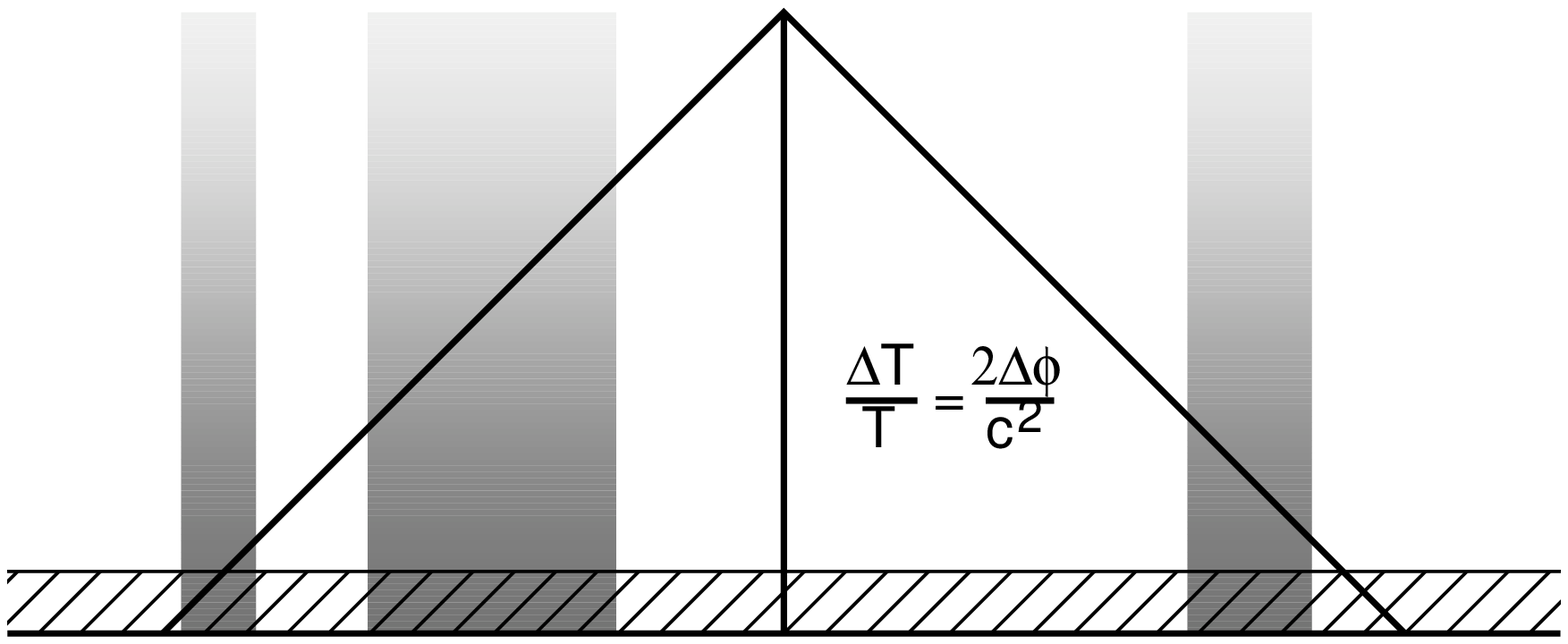}
\vskip 0pt \caption{
Fading potentials cause large-scale anisotropy correlated
with LSS due to the late integrated Sachs-Wolfe
effect.\label{fig:ISW}}
\end{figure*}

\begin{figure*}[b!]
\includegraphics[width=1.00\columnwidth,angle=0,clip]{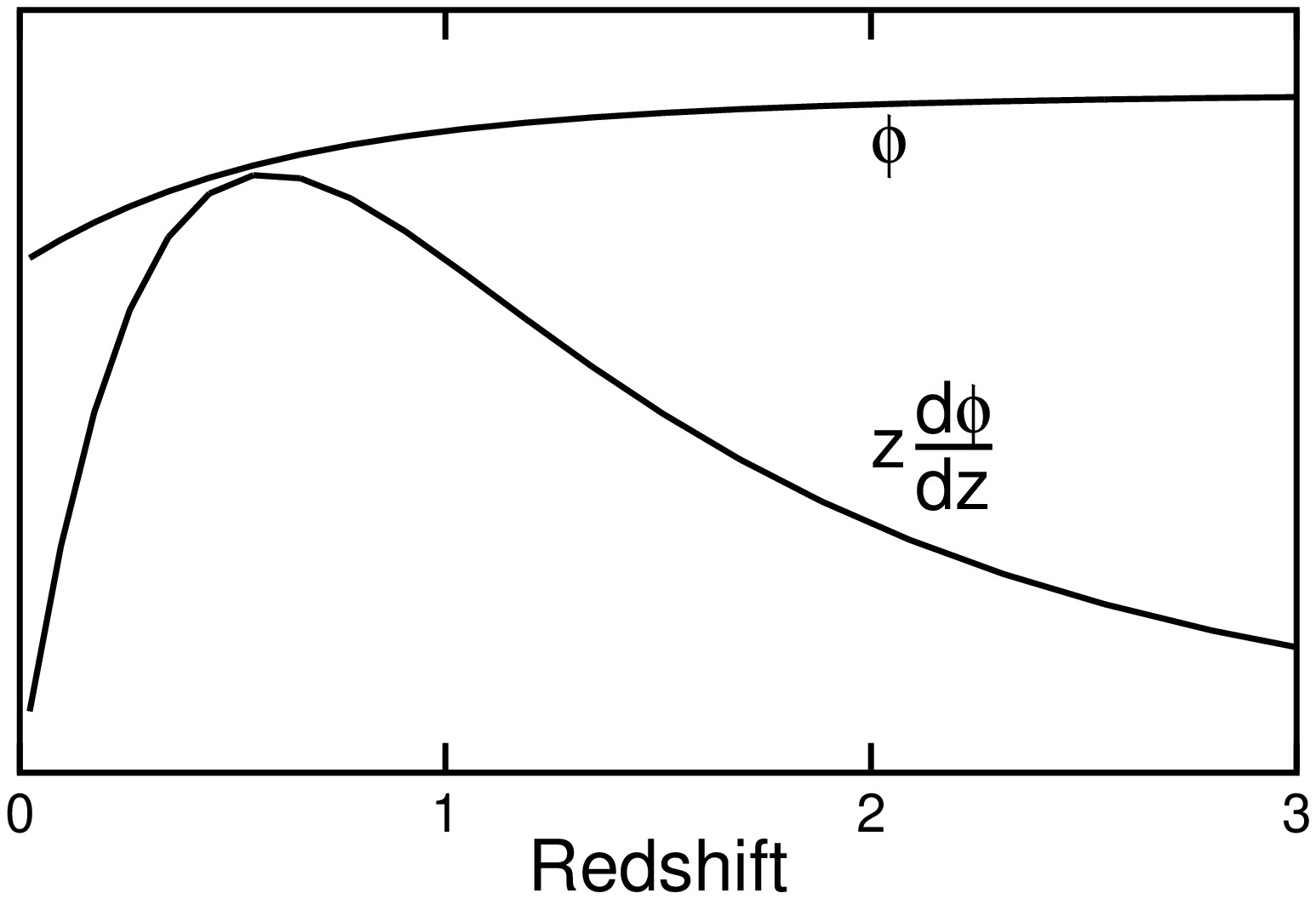}
\vskip 0pt \caption{
Change of potential \vs\ redshift in a $\Lambda$CDM model.
Note that the most significant changes occur near $z = 0.6$.
\label{fig:Dphi-dz}}
\end{figure*}

Two examples of flat ($\Omega_{tot} =1$) models with equal power on all
scales ($n = 1$), plotted on the pre-{\it WMAP} data set, are
shown in Figure \ref{fig:CMB-19Nov02}.  
Both these models are acceptable fits, but the $\Lambda$CDM
model is somewhat favored based on the positions of the
peaks.  The rise in $C_\ell$
at low $\ell$ for the $\Lambda$CDM model is caused by the
late integrated Sachs-Wolfe effect, which is  due to the changing
potential that occurs for $z < 1$ in this model.
The potential changes because the density contrast stops
growing when $\Lambda$ dominates while the Universe
continues to expand at an accelerating rate.  The potential
change during a photon's passage through a structure
produces a temperature change given by $\Delta T/T = 2\Delta\phi/c^2$ 
(Fig. \ref{fig:ISW}).
The factor of 2 is the same factor of 2 that enters into
the gravitational deflection of starlight by the Sun.
The effect should be correlated with LSS that
we can see at $z \approx 0.6$.  Boughn \& Crittenden (2003)
have looked for this correlation using {\it COBE} maps compared
to radio source count maps from the NVSS, and Boughn, Crittenden,
\& Koehrsen (2002) have looked at the correlation of {\it COBE} and
the X-ray background.  As of now the 
correlation has not been seen, which is an area of concern for
$\Lambda$CDM, since the (non)correlation implies
$\Omega_\Lambda = 0 \pm 0.33$ with roughly Gaussian errors.
This correlation should arise primarily from
redshifts near $z= 0.6$, as shown in Figure \ref{fig:Dphi-dz}.
The coming availability of LSS maps based on
deep all-sky infrared surveys (Maller 2003)  should allow a better
search for this correlation.

In addition to the late integrated Sachs-Wolfe effect from
$\Lambda$, reionization should also enhance $C_\ell$ at low
$\ell$, as would an admixture of tensor waves.  Since $\Lambda$,
$\tau_{ri}$ and $T/S$ all increase $C_\ell$ at low $\ell$, and this
increase is not seen, one has an upper limit on a weighted sum of all
these parameters.  If $\Lambda$ is finally detected by the correlation
between improved CMB and LSS maps, or if 
a substantial $\tau_{ri}$, such as the $\tau = 0.1$ predicted by Cen (2003), 
is detected by the correlation between the $E$-mode polarization and
the anisotropy (Zaldarriaga 2003), 
then one gets a greatly strengthened limit on tensor 
waves.

\section{Discussion}

The observed anisotropy of the CMB has an angular power spectrum that
is in excellent agreement with the predictions of the $\Lambda$CDM
model.  But the CMB angular power spectrum is also consistent with
an Einstein-de Sitter model having $\Omega_m = 1$ and a low value
of $H_0 \approx 40$ \kmsmpc.  
The observed lack of the expected correlation
between the CMB and LSS due to the late integrated
Sachs-Wolfe effect in $\Lambda$CDM slightly favors the
$\Omega_m = 1$ ``super Sandage''
CDM model (sSCDM), which, like $\Lambda$CDM, 
is also consistent with the shape of the matter
power spectrum $P(k)$ and the baryon fraction in clusters of galaxies.  But
sSCDM disagrees with the actual measurements of $H_0$ and with
the supernova data for an accelerating Universe.
Thus, $\Lambda$CDM is the overall best fit, but further efforts  to confirm
the CMB-LSS correlation should be encouraged.

\section{Conclusions}

The pre-{\it WMAP} CMB angular power spectrum assembled from multiple 
experiments is very well fit by a six-parameter model.  Of these six 
parameters, the
vacuum energy density $\Omega_V$ and the redshift of reionization
$z_{ri}$ are still poorly determined from CMB data alone.  However,
the well-determined parameters either match independent determinations
or the expectations from inflation:
\bi
\item The baryon density $\omega_b$ is determined to 10\% and agrees with
the value from Big Bang nucleosynthesis.
\item The age of the Universe is determined to 11\% and agrees with
determinations from white dwarf cooling (Rich 2003), main sequence
turnoffs, and radioactive decay.
\item The predicted shape of the LSS power spectrum $P(k)$ agrees with the
observed shape.
\item The curvature $\Omega_k$ is determined to 4\% and agrees with the
expected value from inflation.
\item The spectral index $n$ is determined to 4\% and agrees with the expected
value from inflation.
\ei
The angular power spectrum of the CMB can be computed using well-understood
physics and linear perturbation theory.  The current data set agrees with
the predictions of inflation happening less than 1 picosecond after
the Big Bang, the observations of light isotope abundances from the first
three minutes after the Big Bang, and the observations of LSS
in the current Universe.  The inflationary scenario and the hot
Big Bang model appear to be solidly based on confirmed quantitative
predictions.

The greatly improved CMB data expected from {\it WMAP}, and later {\it Planck},
 should dramatically improve our knowledge of the Universe.

\end{document}